\documentclass[aps,prb,twocolumn,groupedaddress]{revtex4-1}
\usepackage{graphicx} 
\usepackage[dvipdfm]{hyperref}
\hypersetup{colorlinks=true,linkcolor=blue,
  implicit=true,breaklinks=true,pagebackref=true,backref=true,
bookmarks=true,bookmarksnumbered=true,hyperfootnotes=true,debug=true,
  naturalnames=false,citecolor=blue,pdfview=FitH,pdfstartview=FitH,hyperindex=true}
\usepackage{amsmath}
\usepackage{amssymb}
\usepackage{ifsym}
\begin{document}

\title{Filamentary superconductivity across the phase diagram of Ba(Fe,Co)$_2$As$_2$}

\author{H. Xiao$^{1,2\star}$}
\author{T. Hu$^{2,3}$}
\author{S. K. He$^{1}$}
\author{B. Shen$^{1}$}
\author{W. J. Zhang$^{1}$}
\author{B. Xu$^{1}$}
\author{K. F. He$^{1}$}
\author{J. Han$^{1}$}
\author{Y. P. Singh$^{2}$}
\author{H. H. Wen$^{4}$}
\author{X. G. Qiu$^{1}$}
\author{C. Panagopoulos$^{5,6}$}
\author{C. C. Almasan$^{2}$}

\affiliation{$^{1}$ Beijing National Laboratory for Condensed Matter Physics, Institute of Physics, Chinese Academy of Sciences, Beijing 100190, China}
\affiliation{$^{2}$ Department of Physics, Kent State University, Kent, Ohio, 44242, USA}
\affiliation{$^{3}$ State Key Laboratory of Functional Materials for Informatics, Shanghai Institute of Microsystem and Information Technology, Chinese Academy of Sciences, Shanghai 200050, China}
\affiliation{$^{4}$ Department of Physics, Nanjing University, Nanjing 210093, China}
\affiliation{$^{5}$ Division of Physics and Applied Physics, Nanyang Technological University, 637371, Singapore}
\affiliation{$^{6}$ Department of Physics, University of Crete and FORTH, 71003, Heraklion, Greece}

\date{\today}
\begin{abstract}
We show magnetotransport results on Ba(Fe$_{1-x}$Co$_x$)$_2$As$_2$ ($0.0 \leq x \leq 0.13$) single crystals.
We identify the low temperature resistance step at 23 K in the parent compound with the onset of filamentary superconductivity (FLSC), which is suppressed by an applied magnetic field in a similar manner to the suppression of bulk superconductivity (SC) in doped samples.
FLSC is found to persist across the phase diagram until the long range antiferromagnetic order is completely suppressed.
A significant suppression of FLSC occurs for $0.02<x<0.04$, the doping concentration where bulk SC emerges. 
Based on these results and the recent report of 
an electronic anisotropy maximum for 0.02 $\leq x \leq$ 0.04 [Science 329, 824 (2010)], we speculate that, besides spin fluctuations, orbital fluctuations may also play an important role in the emergence of SC in iron-based superconductors.
\end{abstract}


\maketitle
\subsection{INTRODUCTION}
The origin and mechanism of unconventional superconductivity (SC) remain central issues in modern condensed matter physics. Similarly to copper-oxide and heavy-fermion superconductors, the unconventional SC in iron-based superconductors arises in close proximity to antiferromagnetism (AFM). The parent compound of the iron-based superconductors is an AFM metal. Upon doping or applying pressure, the AFM order is suppressed and gives way to SC.  However, unlike the copper-oxide and heavy-fermion superconductors, which mostly are $d$-wave superconductors, the pairing symmetry in this new class of superconductors is more complicated. It can be either $s_{++}$, \cite{JPSJ.79.014710} or $s_{\pm}$.\cite{PhysRevLett.101.057003} There is also evidence for nodal superconductivity.\cite{PhysRevLett.104.087005} Therefore, the origin of iron pnictide SC is still of intense debate. 

Some studies have shown the importance of spin fluctuations, while others have shown the importance of orbital fluctuations to the emergence of SC. Specifically, on one hand, experimental and theoretical studies show evidence that the electron pairing is mediated by antiferromagnetic spin fluctuations \cite{PhysRevLett.101.057003, Basov} and that SC nucleates at antiphase domain boundaries suggesting that spin fluctuations play an important role in the emergence of SC.\cite{XiaoPRB_Ca122}
On the other hand, there is evidence that the multiorbital band structure near the Fermi energy in iron pnictides could enhance orbital fluctuations that become important to the emergence of SC.\cite{Shimojima29042011, PhysRevLett.104.157001}   
Particularly, it has been shown that the uneven occupation of the $d$ orbitals makes the orthorhombic crystal structure more energetically favorable, thus inducing a structural phase transition at $T_s$.\cite{PhysRevB.80.224506, PhysRevLett.101.057003} However, the large electronic anisotropy revealed in  Ba(Fe$_{1-x}$Co$_x$)$_2$As$_2$ precisely where the crystal's C$_4$ rotational symmetry is broken can not be explained based on the  1\% lattice distortion at $T_s$.\cite{Chu_Science} In fact, it has been shown the presence of a C$_4$  structural to C$_2$ electronic symmetry transition in the quasi-particle interference maps of Ca(Fe$_{1-x}$Co$_x$)$_2$As$_2$.\cite{Science_Ca(FeCo)2As2} This electronic symmetry transition could be a result 
of orbital ordering due to the above mentioned inequivalent occupation of the $d_{xz}$ and $d_{yz}$ orbitals.\cite{Chu_Science, PhysRevB.79.054504, PhysRevLett.103.176101, Science_Ca(FeCo)2As2} 

Based on these results, we found that it is imperative to perform measurements that could simultaneously reveal  the effect of spin and orbital fluctuations on the emergence of SC in iron-based superconductors. 
In this paper, we reveal through magnetoresistivity studies the presence of filamentary superconductivity (FLSC) over a wide range of the phase diagram of Ba(Fe$_{1-x}$Co$_x$)$_2$As$_2$  ($0 \leq x \leq 0.13$), from the parent AFM compound to the optimally-doped region, and its disappearance in the over-doped region where the AFM order is suppressed. We show that FLSC, which might have the same origin as the bulk SC, coexists with AFM as a result of competing SC and AFM orders. According to our previous work,\cite{XiaoPRB_Ca122} FLSC and AFM fluctuations are closely correlated. The suppression of  the temperature $T_{fl}$ where FLSC sets in is over the same doping range where a maximum in the orbital order emerges. This  might suggest that the orbital and SC orders are also competing orders. Hence, we speculate that both spin and orbital fluctuations may play important roles for the occurrence of SC in iron-based superconductors.  

\subsection{EXPERIMENTAL DETAILS}
High quality  single crystals of Ba(Fe$_{1-x}$Co$_x$)$_2$As$_2$ were grown using the FeAs flux method.\cite{Ni_PhysRevB.78.214515}  Powder X-ray diffraction measurements were done and the amount of impurities is below the sensitivity of the machine. Typical dimensions of the single crystals are $2\times$0.5$\times$0.06 mm$^{3}$. The in-plane resistivity $\rho$ was measured using the electrical contact configuration of the flux transformer geometry \cite{PhysRevB.55.R3390} and multiple electrodes were fabricated on each sample by bonding Au wires to the crystal with H20E epoxy paste.  The current $I$ was applied in the $ab-$plane and the magnetic field $H$ (up to 14 T) was applied parallel to the $c$ axis of the crystals.

\subsection{RESULTS AND DISCUSSION}
\begin{figure}
\centering
\includegraphics[width=1.0\linewidth]{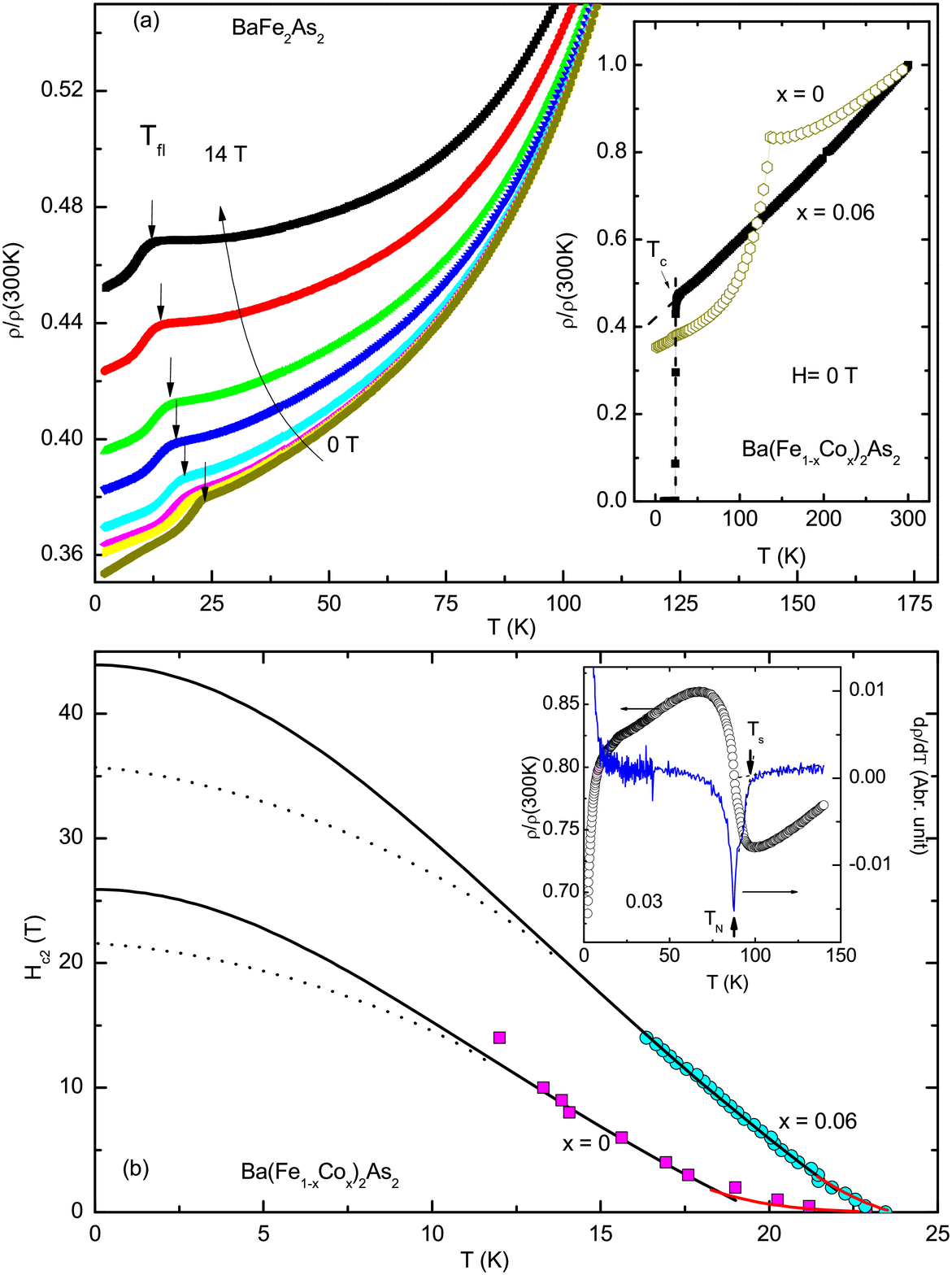}
\caption{\label{fig:Co0and0.06} (Color online)(a) Temperature $T$ dependence of the reduced resistivity  $\rho/\rho(300K)$ curves for  BeFe$_2$As$_2$ measured in a current of 1 mA and different applied magnetic fields; i.e., $H=$ 0, 0.5, 1, 2, 4, 6, 10, 14 T ($H \parallel c$). Inset: Zero-field resistivity curves of Ba(Fe$_{1-x}$Co$_x$)$_2$As$_2$ ($x=$ 0 and 0.06) measured over a wide temperature range (2 to 300 K). (b)  $H-T$ phase diagrams of Ba(Fe$_{1-x}$Co$_x$)$_2$As$_2$ ($x=$ 0 and 0.06). The black solid curves are fits of the data to the GL expression for the upper critical field; the dotted lines are fits with the WHH relationship. The red solid curves are guides to the eye. Inset: Reduced resistivity $\rho/\rho(300K)$ curve and the derivative curve $d\rho/dT$, for $x=$ 0.03 sample. }
\end{figure}

The samples studied in this work have an actual Co concentration $0.0 \leq x \leq 0.13$. The values of the overdoped samples were determined based on the value of the resistive $T_c$.  To determine the actual Co concentration for all the other  samples we extracted the structural $T_s$ and antiferromagnetic $T_N$ phase transition temperatures  from the derivative of the resistivity curve [inset to Fig. \ref{fig:Co0and0.06}(b)]. Specifically, upon cooling from high temperatures $d\rho/dT$ shows a change in slope, followed by a sharp dip as indicated by the arrows. We used these two features to determine $T_s$ and $T_N$, respectively, in correlation to neutron and X-ray measurements. \cite{PRattPhysRevLett.103.087001}  Knowing these transition temperatures for each sample, we determined the actual $x$ value from the temperature - doping ($T-x$) phase diagram of Fig. 3 (Ref. \cite{Ni_PhysRevB.82.024519}). 

Figure \ref{fig:Co0and0.06}(a) depicts the reduced resistivity $\rho/\rho$(300 K) data for the parent compound (no Co doping) measured up to 14 T, while its inset  shows the zero-field data measured over a wide temperature range (2 to 300 K). Notice the presence of a small step (decrease) in the data of the main panel which shifts to lower temperatures with increasing magnetic field (see arrows). Based on a recent report on undoped antiferromagnetic CaFe$_2$As$_2$, \cite{XiaoPRB_Ca122} we identify  the temperature $T_{fl}$ ($T_{fl}= 23.5$ K in the zero-field resistivity data) of this step in resistivity with the onset of filamentary superconductivity. 

Figure \ref{fig:Co0and0.06}(b) shows the $H-T$ phase diagrams generated from the field dependence  of $T_{fl}$ of the $x=0$ sample (magenta squares) and an almost optimally-doped ($x=$ 0.06) single crystal with bulk $T_c$ = 25.1 K (green circles). [Its zero-field-resistivity data are shown in the inset to Fig. \ref{fig:Co0and0.06}(a).] 
A comparison of these $H-T$ phase diagrams shows the striking similarity between them, with the zero-field $T_{fl}$ ($x=0$) and $T_c$ ($x=0.06$) within two degrees of each other, and their similar suppression by magnetic field.
The $H-T$ data of the undoped and optimal doped samples can scale together (data not shown), which might suggest that FLSC and bulk SC have a common origin. Linear fits of these two sets of data give upper critical field $H_{c2}(0)=$ 30.4 T, $T_c=$ 19.5 K, and a slope of - 1.56 T/K for the $x=0$ sample, and $H_{c2}=$ 50.3 T, $T_{c}=$ 22.9 K, and a slope of -2.2 T/K for the $x=0.06$ sample. A fit with the Ginzburg-Landau (GL) expression for the upper critical field, $H_{c2}(T)=H_{c2} (0) [1-(T/T_c)^2]/[1+(T/T_c)^2]$ yields $H_{c2}(0)=$ 25.9 T and $T_{c}=$ 19.7 K for the $x=0$ sample, and $H_{c2}(0)$ = 43.9 T and $T_{c}$ = 22.9 K for the $x=0.06$ sample. The dashed curve is a fit to the Werthamer-Helfand-Hohenberg (WHH) relation, $H_{c2}(0)=-0.7T_c(dH_{c2}/dT_c)$, which gives $H_{c2}(0)= 21.3$ T for the $x=0$ sample, and $H_{c2}(0)$ = 35.3 T for the $x=0.06$ sample. The red solid curves show the high temperature tails for both dopings, typical in iron-based superconductors.

It has been argued that the filamentary SC reported in CaFe$_2$As$_2$ suggests  that even the nominally pure stoichiometric material can spontaneously become electronically inhomogeneous at the nanoscale.\cite{XiaoPRB_Ca122} Such an electronic inhomogeneity at the nanoscale  occurs spontaneously in a nominally uniform system due to competing interactions or competing orders (see Ref. \cite{PhysRevB.79.165122} and references therein).  Inhomogeneity is favorable for a subdominant order to emerge locally, in regions where the competing dominant order vanishes. For example, AFM order can be present inside the vortex core \cite{Hu2012PRL} and FLSC nucleates at the domain walls in the antiferromagnetically ordered parent compound.\cite{XiaoPRB_Ca122} The very similar $H-T$ phase diagrams of undoped BaFe$_2$As$_2$ and undoped antiferromagnetic CaFe$_2$As$_2$  \cite{XiaoPRB_Ca122} show a common signature for FLSC in these parent compounds and, therefore, imply that the AFM order is also a competing order to superconductivity in the BaFe$_2$As$_2$ system.

We note that the step in the low temperature resistivity [see Fig. 1(a)] is not always observed in the undoped samples.\cite{ColombierPhysRevB.79.224518, DucanHighpressureBa122, IshikawaPressureBa122, Ni_PhysRevB.78.214515} For example, Tanatar et al., studied five different samples of undoped BaFe$_2$As$_2$ and found that only two samples showed the partial SC transition. \cite{TanatarBa122} We also made measurements on  a number of single crystals of the parent compound and confirmed that this step is observed only for small and thin single crystals, (with a thickness of $\sim$ 50-60 $\mu$m) with shiny surfaces (after careful cleaving), whereas in the resistivity of larger samples, which contain a number of smaller crystals, this feature is not there. High quality single crystals are, therefore, better for the detection of filamentary superconductivity. A similar conclusion has been reached by Park et al., in their study of textured SC in CeRhIn$_5$.\cite{Park2012} 

\begin{figure}
\centering
\includegraphics[trim=0cm 0cm 0cm 0cm, clip=true, width=0.5\textwidth]{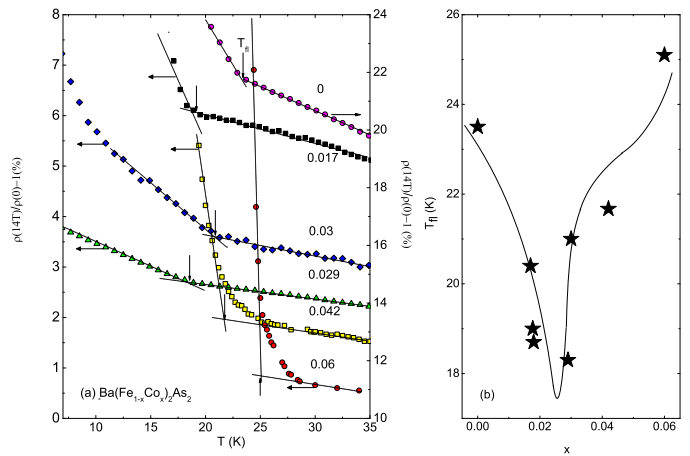}
\caption{\label{fig:magnetoresistance} (Color online) (a) Temperature $T$ dependence of magnetoresistivity $\rho(14$ T$)/\rho(0)$-1 for the $x=$ 0, 0.017, 0.029, 0.03, 0.042, 0.06 and 0.08 samples. The vertical arrows mark the temperature $T_{fl}$ where filamentary superconductivity emerges. (b) Doping $x$ dependence of $T_{fl}$.}
\end{figure}

Since the step in resistivity is sometimes hard to detect for reasons discussed above, we performed magnetoresistivity (MR) measurements, which is a more sensitive method to detect the small decrease in resistivity due to FLSC.  To determine how FLSC  evolves across the phase diagram of  Ba(Fe,Co)$_2$As$_2$, we measured the temperature dependence of the magnetoresistivity $\Delta \rho/\rho(0) \equiv [\rho(14 $T$)-\rho(0)]/\rho(0)$ of single crystals with different Co doping  [Fig. \ref{fig:magnetoresistance}(a)]. 
 With decreasing temperature, the magnetoresistivity first increases almost linearly, then shows a sudden deviation (kink) at a certain temperature. 
We associate this sudden change in magnetoresistivity with the appearance of FLSC and identify its temperature as $T_{fl}$ for all the samples studied in this work since we confirmed that the temperature of this kink in the magnetoresistivity of the parent compound coincides with $T_{fl}$ and, in addition, FLSC, like SC, is strongly affected by magnetic field. We summarize the doping dependence of $T_{fl}$ in Fig. 2(b) and Fig. 3.

It is worth to mention that we compared samples with the same Co concentration but different thicknesses and found that as the thickness decreases the signature of FLSC (step in resistivity) is more pronounced. However, the kink in the MR data is at the same $T_{fl}$, for all samples (regardless their thickness), even for the thick samples  that do not show any low temperature step in their resistivity. Furthermore, the magnitude of MR when approaching $T_{fl}$ from above is the same regardless the thickness of the sample. Therefore, we conclude that, even though the low temperature step in resistivity is not always detectable, FLSC exists in all samples and can be better detected through the more sensitive MR measurement.

The plot of Fig. 2(b) shows that, in fact, $T_{fl}$ displays a non-monotonic dependence on Co doping, with a minimum for $x\approx$ 0.02 to 0.03. Interestingly, Jiun-Haw Chu et al. \cite{Chu_Science} have recently reported a maximum electronic nematic order over the same range of Co concentrations, around  the onset of bulk superconductivity ($x=$ 0.03). 
 This suggests a maximum orbital order around the onset of bulk SC. 
The fact that $T_{fl}$ of FLSC shows a minimum over this same doping range may suggest that orbital  and SC orders are also competing orders. So orbital fluctuations could also contribute to the pairing mechanism in iron-based superconductors. Nevertheless, since, as discussed in our previous work,\cite{XiaoPRB_Ca122} FLSC nucleates at the domain walls, a magnetic origin of SC is an important ingredient. So spin and orbital degrees of freedom might be strongly coupled to each other. We speculate that both spin and orbital fluctuations might contribute to the pairing mechanism of unconventional superconductivity in iron-based superconductors. 

\begin{figure}
\centering
\includegraphics[trim=0cm 0cm 0cm 0cm, clip=true, width=0.5\textwidth]{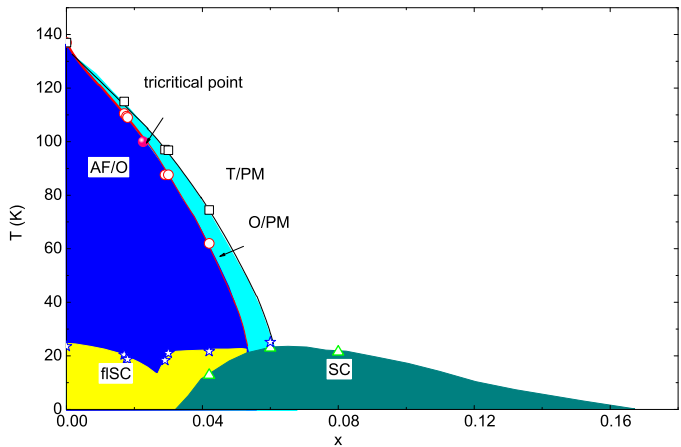}
\caption{\label{fig:Phasediagram} (Color online) Temperature $T$ vs. doping $x$ phase diagram of Ba(Fe$_{1-x}$Co$_x$)$_2$As$_2$ single crystals. The phase boundaries are taken from Ref. \cite{Ni_PhysRevB.82.024519}. The open symbols represent data from this work: structural  $T_s$ (squares), antiferromagnetic $T_N$ (circles), filamentary superconductivity $T_{fl}$ (stars), and
bulk superconductivity $T_c$ (triangles) transition temperature determined from magnetization measurements (data not shown).  The red and black solid lines denote first-order and second-order transitions, respectively (taken from Ref. \cite{PhysRevB.83.134522}). The pink solid circle is the magnetic tricritical point, $x^{m}_{tr}=$ 0.022. }
\end{figure}

Figure \ref{fig:Phasediagram} depicts a composite plot of the phase diagram of the doping dependence of $T_s$, $T_N$, $T_c$ and $T_{fl}$.
$T_s$ and $T_N$ nearly coincide at $x=0$ but they are slightly different for the underdoped samples. Increasing doping suppresses the AFM order, gradually giving way to superconductivity. Bulk superconductivity (triangles, determined from magnetization measurements, data not shown) emerges at $x\approx$ 0.03 and $T_c$ attains a maximum value for  $x \approx$ 0.06, where the zero field magnetic and SC transition temperatures are equal. Compared with the bulk $T_c$, $T_{fl}$ (stars) is less doping dependent, FLSC is already present in the undoped, parent compound and persists up to about $x=$ 0.06, where long range AFM order vanishes. This latter fact suggests again that there is a close relationship between FLSC and AFM. In fact, previous experiments on CaFe$_2$As$_2$ showed that FLSC nucleates at the AFM domain walls, \cite{XiaoPRB_Ca122} i.e., where the AFM exchange interaction is suppressed and the local AFM fluctuations are enhanced.  This would lead to FLSC with $T_{fl}$ close to optimum $T_c$ and slightly doping dependent. Notably, the coexistence of antiferromagnetism and FLSC is inhomogeneous. It is, therefore, plausible that the increase in $T_c$ is, indeed, a result of the suppression of AFM, with optimum $T_c$ emerging where the AFM fluctuations are enhanced. When the AFM transition temperature $T_N$ is suppressed to zero, one may anticipate the presence of a quantum critical point.\cite{PhysRevB.79.014506}  All these results and previous reports of FLSC in other iron-based superconductors such as SrFe$_2$As$_2$,\cite{PhysRevLett.103.037005} or other systems, such as heavy fermion superconductors and cuprate superconductors,\cite{Park2012, CeRhIn5_flSC_Pressure,PetrovicEPL2001, PRB_URu2Si2, 
PRL_stripeorder_Ba_LCO} are consistent with the picture that an inhomogeneous FLSC appears prior to the occurrence of homogeneous bulk SC. So, inhomogeneous SC could be a general feature of unconventional superconductors and it is present only when long range AFM order exists, hence, it is a natural result of competing orders.

\subsection{SUMMARY} 
In summary, we show that FLSC, which might have the same origin as the bulk SC, coexists with AFM as a result of competing SC and AFM orders. The temperature $T_{fl}$ where FLSC sets in is close to the optimal bulk transition temperature $T_c$, suggesting that FLSC is a precursor state to bulk SC. Since the suppression of $T_{fl}$ is in the vicinity of the onset of bulk superconductivity ($x\approx$ 0.03), in the doping range where a maximum in the orbital order emerges, we speculate that besides spin fluctuations, orbital fluctuations may also play an important role in the emergence of unconventional SC in iron-based superconductors.

\subsection{ACKNOWLEDGMENTS} 
We acknowledge financial support by NSFC, project no. 11104335 and the MOST of China, project nos. 2011CBA00107, 2012CB921302. The work at  Kent State University  was supported by the National Science Foundation (grant NSF DMR-1006606 and DMR-0844115) and by the ICAM Branches Cost Sharing Fund from Institute for Complex Adaptive Matter. CP acknowledges support from MEXT-CT-2006-039047, EURYI, National Research Foundation, Singapore.
\\
$^{\star}$Corresponding author; hxiao@iphy.ac.cn.

\end{document}